\documentclass[12pt]{article}
\usepackage[margin=1in]{geometry}

\usepackage{moreverb}
\usepackage{pdfpages}
\usepackage[utf8]{inputenc}
\usepackage{amsmath}
\usepackage{amsfonts}
\usepackage{amssymb}
\usepackage{tikz}
\usepackage{url}
\usepackage{algorithm}
\usepackage{algpseudocode}
\usepackage{varwidth}
\usetikzlibrary{bayesnet}
\usetikzlibrary{arrows}
\usepackage{color}
\usepackage{graphicx}
\usepackage{subcaption}
\usepackage{lipsum}
\usepackage{caption}
\DeclareCaptionFont{10pt}{\fontsize{10pt}{10pt}\selectfont}
\usepackage{float}
\usepackage[section]{placeins}

\usepackage{natbib}
\setcitestyle{authoryear,open={(},close={)}} 

\title{Detecting Changes in the Transmission Rate of a Stochastic Epidemic Model}

\author{
  Jenny Huang, 
  Raphaël Morsomme, 
  David Dunson, and
  Jason Xu \\
  Department of Statistical Science, Duke University
}
\date{}

\begin{document}
\maketitle

\begin{abstract}Throughout the course of an epidemic, the rate at which disease spreads varies with behavioral changes, the emergence of new disease variants, and the introduction of mitigation policies. Estimating such changes in transmission rates can help us better model and predict the dynamics of an epidemic, and provide insight into the efficacy of control and intervention strategies. We present a method for likelihood-based estimation of parameters in the stochastic SIR model under a time-inhomogeneous transmission rate comprised of piecewise constant components. In doing so, our method simultaneously learns change points in the transmission rate via a Markov chain Monte Carlo algorithm. The method targets the exact model posterior in a difficult missing data setting given only partially observed case counts over time. We validate  performance on simulated data before applying our approach to data from an Ebola outbreak in Western Africa and COVID-19 outbreak on a university campus. \end{abstract}

\bigskip \textit{Keywords}: stochastic epidemic models; data-augmented MCMC; Bayesian change point detection; incidence data; COVID-19; likelihood-based inference

\newpage 

\section{Introduction}\label{sec1}

\label{sec:intro}
This article considers the statistical task of estimating epidemic parameters from observational data when the rate of disease transmission varies over time. In particular, we seek to model the transmission rate under an SIR model so that it can flexibly capture changes over time, yet remains parsimonious enough to retain tractable inference and to avoid overfitting the available data.

Recent outbreaks have been characterized by changes in social distancing behaviors and economic policies, control or mitigation measures, the emergence of new disease variants, and many other factors which may lead to changes in the disease transmission rate \citep{zelner2021accounting, eikenberry2020mask}. The common assumption that the transmission rate is fixed over time may no longer be appropriate in such cases, and classical SIR models with a constant transmission parameter, denoted $\beta$, often struggle to fit observed trends in data from recent epidemics such as COVID-19 and Ebola. As a result, several studies have posited time-varying rates, relaxing the assumption of a constant per-contact infection rate made in classical Susceptible-Infected-Removed (SIR) models. Many of these approaches operate within deterministic models \citep{dehning2020inferring} \citep{hong2020estimation} and their variants. However, deterministic models do not lend themselves to natural probabilistic interpretations, so that resulting estimates do not come equipped with measures of uncertainty. Even when an observation or sampling model such as Gaussian or binomial emissions is employed on top of a deterministic mean model, the uncertainty quantification should only be interpreted as a reflection of the measurement error rather than the underlying stochasticity of the model. In order to account for uncertainty more accurately and interpretably, we focus on the stochastic version of the SIR model.

In the stochastic epidemic model, the spread of disease is described according to a probability law rather than a system of differential equations \citep{allen2006stochastic}. The canonical formulation specifies that the disease moves through the population as a Markov jump processes (MJP), which agrees with the deterministic formulation in the large-population limit \citep{andersson2012stochastic}. Unlike diffusion approximations for large populations \citep{dureau2013capturing,fintzi2021linear}, the Markov jump process formulation operates on a discrete state space and is appropriate in early as well as later stages of an outbreak. These \textit{stochastic} epidemic models (SEM) form a class of compartmental models of disease that replace deterministic transition rates by instantaneous jump probabilities, and as a result evolve as continuous-time stochastic processes. When it comes to learning how the rate of transmission in such models changes over time, a much smaller literature has explored inference under the general stochastic epidemic model. 

One reason this space is relatively  underexplored stems from  computational difficulties that arise when the data are only partially informative. Epidemic data are commonly reported as incomplete summaries of a process that evolves continuously through time. For instance, a common setting in observational studies provides \textit{incidence data}, comprised of new infection counts given on some fixed schedule (e.g. weekly, monthly). With incidence data, however, the exact infection and removal times are unknown. In this paper, we will focus on this data setting where only the number of new infections at a discrete set of times are observed.
In this partially-observed  setting, evaluating the marginal likelihood of the observed data under a stochastic epidemic model requires access to the finite-time transition probabilities of the process \citep{guttorp2018stochastic}. These quantities are difficult to compute in our model, requiring integration over the high-dimensional space of all configurations of the trajectories between observed times consistent with our observed data \citep{boys2007bayesian}. Although integrating over this space is possible in principle \citep{hobolth2009simulation,ho2018birth}, available methods are computationally intensive and often impractical beyond the most simple cases \citep{kypraios2018bayesian}. As a result, methods to address fitting stochastic epidemic models to partially-observed data have been a subject of research over the last fifteen years, and recent advances, particularly in data augmentation and approximation methods, have started to overcome these challenges in inference \citep{xu2016bayesian}\citep{fintzi2021linear}. 

In recent attempts to introduce change points in the stochastic epidemic model, some studies have taken a fairly rigid approach by fixing change point locations based on knowledge of the exact dates on which policy changes occurred \citep{lekone2006statistical} \citep{ward2021individual}. Unfortunately, this rigid approach can lead to biased results, failing to account for lag times and ruling out the possibility of other change points that are unaccounted for. Here, we stress the importance of learning change points flexibly from data. While we place this task in the framework of change points, classical change point detection algorithms are poorly suited when the goal is to detect changes in the \textit{transmission rate} under a model-based approach. The transmission rate is related to the observed time series data through the mechanistic model.
In particular, the overall infection rate is already varying over time, given by $\beta S(t) I(t)$. In replacing the constant transmission rate $\beta$ by a function $\beta(t)$, one should seek formulations that balance more flexibility with enough parsimony so that it does not overwhelm the model-based dynamics defined via  interactions between $S(t)$ and $I(t)$. 

To this end, our paper seeks to capture changes in the per-contact transmission rate by replacing the constant force with a piecewise constant function $\beta(t)$, learning where to best allow change points in its segments from the data. 
There is a mature body of literature focused on multiple change point detection \citep{barry1993bayesian,chib1998estimation,davis2006structural,fearnhead2006exact}. Frequentist methods commonly rely on a test statistic, such as the likelihood ratio or CUSUM statistic, to test for changes in the distribution of the data, followed by a model selection step, typically using a LASSO type penalty, to determine the number of parameters defining the signal \citep{davis1995testing,cappello2021scalable,fryzlewicz2014wild}. In contrast, Bayesian approaches formulate the multiple change point problem in terms of a sequence of hidden discrete state variables that evolve as a Markov process, and use methods such as MCMC or dynamic programming to estimate the posterior distribution of change points \citep{chib1998estimation,fearnhead2006exact}. The majority of these existing methods concern cases of independent and identically distributed random variables. This is not the case under the SIR model, where event times follow a Poisson process with time-varying intensity function $\beta S(t)I(t)$--- the number of new infections depends on both the transmission rate and the sizes of the disease compartments at a given time. For these reasons, general-purpose change point detection algorithms are ill-equipped for our task within this mechanistic model. Again, we seek to allow heterogeneity in the transmission rate through change points, rather than to detect change points on the sequence of observed counts directly.

To jointly address these considerations, we propose a fully Bayesian inferential framework for learning changes in the transmission rate of the general stochastic epidemic model. In related work on time-varying parameters in epidemic models, \cite{kypraios2018bayesian} fit the force of infection with second-order B-splines using data with exact recovery times and unknown infection times. However, they do so in a non-parametric model that ignores the mass action assumption of the SIR, effectively replacing the intensity $\beta S(t)I(t)$ entirely by a quite flexible function $\beta(t)$. In contrast, we preserve the interaction term between susceptibles and infectious individuals that undergirds the SIR dynamics, learning a rate function $\beta(t) S(t)I(t)$ where $\beta(t)$ is a step function that only features change points when there is sufficient evidence from the data. This leads to a more parsimonious model, while facilitating samplers targeting the posterior distribution of the transmission rate parameters.
Interpretable outputs can then be used to assess the efficacy of proposed public health interventions such as travel restrictions, school closures, and lockdowns.

In terms of Bayesian computation, \cite{boys2007bayesian} and \cite{kypraios2018bayesian} employ reversible jump (RJ) MCMC for posterior computation of time-varying epidemic parameters. The dimension varies under this approach when proposing to increase or decrease the number of steps, necessitating a RJ-MCMC sampler. Efficient dimension changing moves can be notoriously difficult to design, however, and RJ-MCMC algorithms often exhibit slow mixing
 \citep{kypraios2018bayesian}. Here we present an alternate approach more similar in spirit to the hidden Markov multiple change point model \citep{chib1998estimation}, bypassing the need for the reversible jump step.  That is, rather than estimating the sequence of change points directly, we introduce a latent sequence of indicators, $\{\Delta_t\}$, that take the value 1 at change point locations and 0 otherwise. When the length of the observed time series is $T$, the method we propose simply samples a fixed-dimensional sequence of indicators of length $T-1$. To account for the temporal dependence between change point locations, we infer the hidden states of the Markov process $\{\Delta_t\}$. Further leveraging on recent developments in data augmentation (DA) MCMC for stochastic SIR models \citep{morsomme2022uniformly}, we propose a simple Metropolis-within-Gibbs sampling scheme.

The remainder of the paper is structured as follows: section \ref{sec:methods} provides a description of the data, an introduction to the stochastic SIR model, and a detailed description of
our change point model and computational algorithm.
 In section \ref{sec:results} the model is applied to simulated data followed by applications to Ebola and COVID-19 outbreaks with suspected change points. Section \ref{sec:discussion} closes with a discussion and highlights avenues for future work.

\section{Methods}
\label{sec:methods}

\subsection{Data setting and the stochastic SIR model}

In this article, we focus on the \textit{incidence data} setting in which new infection counts are collected at discrete reporting times. This is typical of observational studies of epidemic data, but presents a challenge for exact model-based inference due to the missing information between observation times \citep{fintzi2021linear}. In particular, given a set of observation times $t_{0:K}$, the observed incidence consists of a $K$-dimensional vector $\textbf{I}_{1:k} = (I_1, \dots, I_K)$, where $I_k$ denotes the number of new cases reported during the interval $(t_{k-1},t_k]$.

We are interested in learning the parameters of a stochastic SIR model to these data. This widely used compartmental model  describes the spread of an epidemic through a population divided into three compartments based on disease status: susceptible ($S$), infectious ($I$), and removed ($R$) individuals. Individuals transition from $S$ to $I$ when they become infected and from $I$ to $R$ when they recover or are removed (eg. death). 
While the deterministic version is defined through a system of differential equations, the stochastic SIR model evolves as a continuous-time Markov chain (CTMC) $X = \{ X(t); t > 0\}$ where the components $X(t)=\{S(t), I(t), R(t)\}$ track the respective compartments $S(t)$, $I(t)$ and $R(t)$ denoting the number of susceptible, infected and removed individuals in the population at time $t$\citep{allen2006stochastic} (Figure \ref{diag:SIR}).

\begin{center}
		\begin{tikzpicture}[thick]
			\node[draw,	circle, minimum width=1cm] at (0  ,0) (block1) {S(t)};
			\node[draw,	circle, minimum width=1cm] at (3.5,0) (block2) {I(t)};
			\node[draw,	circle, minimum width=1cm] at (7  ,0) (block3) {R(t)};
			
			\draw[-{[scale=2]}] (block1) -- (block2)
			node[pos=0.45,above]{$\beta S(t)I(t)$};
			\draw[-{[scale=2]}] (block2) -- (block3)
			node[pos=0.45,above]{$\gamma I(t)$};
		\end{tikzpicture}
  \label{diag:SIR}
\end{center}

The population is assumed to be closed and homogeneously mixing, ignoring demographic dynamics such as births and deaths \citep{bailey1975mathematical}. This implies that contacts between individuals follow independent Poisson processes with rate $\beta$, which can be interpreted as the infection rate of the contagious disease. We further assume that the infectious periods follow independent exponential distributions with rate $\gamma$ for exposition. As a result, the stochastic SIR model is a bivariate Markov process that is characterized by the instantaneous transition probabilities below:

\begin{multline*}
P(S(t+h), I(t+h)) = (s+k, i+j) \mid (S(t), I(t)) \\ 
= (s, i)) = \begin{cases} 
   \beta S(t)I(t) + o(h) & (k,j) = (-1, +1) \\
   \gamma I(t) + o(h) & (k,j) = (0, -1) \\
  1 - (\beta S(t)I(t) + \gamma I(t)) + o(h) & (k,j) = (0, 0) \\
  o(h) & \text{otherwise}
\end{cases}
\end{multline*}

\subsection{Time-varying infection rate}

The model as specified above posits that the transmission rate $\beta$ per contact between susceptible and infectious person is constant over time. Here, we relax this standard assumption to allow this rate $\beta(t)$ to vary over time, and choose to model it as a piece-wise constant function, without specifying in advance the points at which its constant values change. This choice keeps the original mass action dynamics of SIR, while allowing for jumps in the per-contact transmission rate. That is, given some $J$ change point locations $(c_1, \dots, c_J) \subset (t_1, \dots, t_K)$,

\begin{align}
    \beta(t) = \beta_k \textbf{  } \forall t \in [c_{k} \leq t \leq c_{k+1}] . 
    \label{piecewise_beta}
\end{align}

Under this piece-wise constant formulation, the points $\{c_k\}$ are naturally interpreted as change point locations. To learn their locations jointly with parameter inference, we introduce a sequence of latent binary variables which denote the presence of such change points. At each observation time $t_k$, the latent indicator variable $\Delta_t$ takes on the value of 1 to indicate the existence of a change in transmission rate and 0 if there is no change. 

\begin{align}
    \label{delta}
    \Delta_t  = \left\{
    \begin{array}{ll}
        1 & \quad \text{if t = $c_k$ for some k} \\
        0 & \quad \text{otherwise}
    \end{array}
\right.
\end{align}
In particular, the locations of the 1's in the sequence $\{\Delta_t\}$ encode the locations of change points $\{c_k\}$ where $\beta(t)$ takes on a new value. Hence, our formulation will reduce the task of inferring a time-varying infection rate $\beta(t)$ to estimating the change point sequence $\{\Delta_t\}$, and performing efficient parameter updates jointly on the locally constant segments.

\subsection{The complete data likelihood}

As a CTMC, the time until the next infection event follows an exponential distribution with rate $\beta(t)S(t)I(t)$, and the time until next removal event follows an exponential distribution with rate $\gamma I(t)$. Considering the epidemic on one of the intervals $(c_k, c_{k+1}]$, we obtain the following likelihood of the SIR model with a piece-wise infection rate

\begin{align}
    \label{complete_data_likelihood_sir}
    \mathcal{L(\mathbf{\theta}; \mathbf{X})} 
    & = \prod_{k} \left( \prod_{j \in \mathcal{I}_k} \beta_k I(\tau_j^I) \prod_{l \in \mathcal{R}_k} \gamma \, \exp\bigg\{-\int_{c_k}^{c_{k+1}} \beta_k I(t)S(t) + \gamma I(t) dt\bigg\} \right) \nonumber \\
    & = \gamma^{n_R} \prod_{k} \left( \beta_k^{n_{I_k}} \prod_{j \in \mathcal{I}_k} I(\tau_{j}^I) \, \exp\bigg\{-\beta_k \int_{c_k}^{c_{k+1}}  I(t)S(t) dt + \gamma \int_{c_k}^{c_{k+1}}  I(t) dt\bigg\} \right)
\end{align}

Here, $\mathbf{X}$ denotes an epidemic trajectory consisting of all infection and removal times, $k$ is the index on the step of the piece-wise transmission rate, and $\{c_k\}$ are the sequence of change point times. The index sets $\mathcal{I}_k= \{j : \tau_j^I \in (c_k, c_{k+1}] \}$ and $\mathcal{R}_k = \{j : \tau_j^R \in (c_k, c_{k+1}] \}$ correspond to the individuals that are infected and removed during the interval $(c_k, c_{k+1}]$. The quantity $n_{I_k} = |\mathcal{I}_k|$ denotes the number of infection and removal events over this same time interval, and $n_R = \sum_k |\mathcal{R}_k|$ denotes the total number of removal events. Note that due to the Markov property, the likelihood over an interval $(c_k, c_{k+1}]$ where the transmission rate is locally constant takes on the same form as the usual stochastic SIR likelihood with constant parameter $\beta_k$ for an outbreak restricted to that interval, and the likelihood of the process on the entire observation period is a product over these intervals. 

\subsection{Model parsimony and promoting sparsity in change points} 
To control model complexity and avoid overfitting, we next seek to impose parsimony on the rate function given by (Eq. \eqref{piecewise_beta}). It is undesirable to assign each $k$th observation interval its unique $\beta_k$ value; the number of model parameters increases with the number of times we observe incidence, and we may expect non-identifiability when $\beta(t)$ is too flexible since $I(t)$ is unknown and the overall infection rate to be learned takes the form $\beta(t) S(t) I(t)$.

An intuitive way to avoid this is by promoting \textit{local constancy}: that is, we would like to encourage runs of constant values of $\beta_k$ for multiple observation intervals, $(t_{k-1}, t_{k}]$, in a row \textit{a priori}. Then, the rate function only changes when sufficiently supported by the data. 
We may naturally encode this through prior probability distributions on the vector of changes $\Delta$. We employ a simple Markov process for how $\Delta$ changes, governed by the transition matrix
\begin{align}
    \Pi =
\begin{bmatrix}
\pi_{00} & \pi_{01}\\
\pi_{10} & \pi_{11}
\end{bmatrix}.
\label{pi_markovchain_transition_matrix}
\end{align}
Here $\pi_{ij}$ denotes the probability of transitioning from state $i$ to $j$ from one time point to the next. 

Note if there is a change point between the previous and current time step (i.e. $\Delta_t = 1$) the probability of another change point occurring at the next time step is $\pi_{11}$, while the probability of the rate remaining constant is represented by the transition into a non-change point state $\pi_{10}$. Likewise, if there was no change point between the previous and current time step (i.e. $\Delta_t = 0$), the probability of a change point at the next time is $\pi_{01}$ while that of remaining in a constant state is $\pi_{00}$. We see that this prior promotes local constancy as long as $\pi_{\cdot 1}$ is small---that is, it encourages sequences $\Delta$ to have runs of zeros \textit{a priori}.  This effect is akin to first-order methods to promote local constancy such as the intuition behind the fused Lasso \citep{tibshirani2005sparsity} when $\pi_{11}=\pi_{01}$, but allows the user to additionally penalize back-to-back change points by setting  $\pi_{11} < \pi_{01}$. Doing so is desirable in settings when rapid changes in policy, behavior, or the emergence of new variants may be unrealistic. The prior for $\Delta$ induces a hidden Markov model that captures the temporal dependence of the infection rate between a two-state latent variable over time  \citep{abraham2022multiple,fox2008hdp}. An analysis of the impact of hyperparameter choices for $\pi$ on change point recovery is included in the Appendix (Figure \ref{fig:change_point_sensitivity}).

While we design priors on $\Pi$ to promote sparsity in change points, note that the Bayesian view of Occam's razor \citep{jefferys1992ockham} automatically protects against over-fitting to an extent. In particular, as the number of change points increases, the size of the effective model increases in turn as it implies more segment-specific parameters $\beta_k$ to estimate. Viewed as a Bayesian model selection problem, the posterior probability on the model corresponding to a particular number of change points is proportional to the prior probability multiplied by a marginal likelihood obtained by integrating out the $\beta_k$ parameters. This marginal likelihood is formed as a product of likelihoods conditional on the $\beta_k$ by their prior, followed by integrating out the $\beta_k$'s. In this sense, the marginal likelihood automatically tends to decrease as the number of effective parameters grow. Hence, the posterior will favor parsimonious models with relatively few change points, only adding change points if they are clearly supported by the data.

This observation is supported empirically, detailed further in our simulation study. On each constant segment of the transmission rate $\beta_k$, we place a gamma prior with weakly informative hyperparameters, $\beta_k \sim Gamma(1, 1)$. Even with these weakly informative priors, we find that the posterior distribution is able to concentrate successfully around the true value (Figure \ref{fig:sim1_diagnostics}), indicating that the data are quite informative of transmission rate within each segment.
 
\subsection{Posterior computation via Metropolis-within-Gibbs}
The model development in the previous sections is designed to admit efficient posterior sampling schemes. 
Here, we derive a data-augmented Markov chain Monte Carlo (DA-MCMC) sampler targeting the posterior distribution of the parameters of interest. Recall the observed incidence data consist of counts of new infections at discrete time points, providing an incomplete view of the continuous-time epidemic process. In particular, the times at which an individual in the population either becomes infected or removed are unavailable. The complete-data likelihood in Eq. \eqref{complete_data_likelihood_sir} is therefore out of reach, and strictly speaking, is related to the likelihood of the observed incidence via marginalizing over all possible sets of missing event times compatible with the observations. This entails an unwieldy integration step over the space of latent complete-data trajectories, for which no closed form expression is available. 

Instead, we model the unobserved event times as latent variables, accounting  for this marginalization step by using a data augmentation approach \citep{tanner1987calculation}. This allows us to conduct inference under the exact model posterior distribution over the parameters and latent variables. Though the MCMC algorithm must explore an expanded space, parameter updates now take advantage of the complete-data likelihood given a configuration of the latent variables that ``complete" the observed data. 

Specifically, we alternate between three types of updates: (i) updates of the transition rate parameters $\Pi$, (ii) joint updates of the change point locations $\Delta$ and steps of the transmission rate $\beta$, and (iii) updates of the latent infection and removal times $X$.

First, we update $\pi_{01}$ and $\pi_{11}$. These parameters are assigned independent beta priors. Because the total number of change points follows a binomial distribution, it follows that the full conditional distribution remains beta distributed. This allows us to update $\Pi$ using a Gibbs sampler (Eq. \eqref{pi_conditional}). 
Specifically, we use the  prior $\pi_{ij} \sim Beta(a_{ij}, b_{ij})$, with expectation $a_{ij}/ (a_{ij} + b_{ij})$ and variance $(a_{ij}b_{ij})/[(a_{ij}+b_{ij})^2(a_{ij}+b_{ij}+1)]$. The conditional distribution for the transition probabilities is then
\begin{align}
    \pi_{01} \mid \Delta \sim Beta(a_{01} + n_{01}, b_{01} + n_{00}),
    \label{pi_conditional}
\end{align}
where $n_{01}$ is the number of transitions in $\Delta$ from state 0 to 1 and and $n_{00}$ is the number of transitions from state 0 to 0. Similarly, $\pi_{11} \mid \Delta \sim Beta(a_{11} + n_{11}, b_{11} + n_{10})$.

Next, the parameters $\{\Delta_t\}$ and $\beta(t)$ are proposed jointly in a Metropolis-Hastings step. Specifically, we propose $\{\Delta_t\}$ as a realization of a Markov chain with transition matrix $\Pi$. The likelihood for the hidden state sequence is 
\begin{align}
    P(\Delta \mid \nu_1, \Pi) = \nu_1 \prod_{i=2}^T \pi_{\Delta_{i-1}\Delta_i} = \pi_{00}^{n_{00}} \cdot \pi_{01}^{n_{01}} \cdot \pi_{10}^{n_{10}} \cdot \pi_{11}^{n_{11}}
\end{align}
where $n_{ij}$ is the number of transitions in $\{\Delta_t\}$ from state $i$ to $j$ and  $\nu_1 = Pr(\Delta_1 = 1)$ is the initial probability, which follows the same distribution as $\pi_{01}$. Within this joint Metropolis-Hastings update, we may now propose $\beta(t)$ segment-by-segment conditional on $\{\Delta_t\}$. Each constant segment of $\beta(t)$ is assigned a gamma prior; the full conditional distribution remains gamma-distributed by conjugacy.  Hence, for each of these segments, we place the prior $\beta_{k} \sim Ga( a_0, b_0 )$, denoting the parametrization with expectation $a_0/b_0$ and variance $a_0/b_0^2$. The conditional distribution of each segment $(\beta_{k} | \Delta, X)$ then follows
\begin{align}
    \beta_{k} \mid \Delta, X \sim Gamma \left( a_0 + n_T, b_0 + \int_{c_k}^{c_{k+1}} I(t)S(t) dt \right). 
    \label{segment_full_conditional}
\end{align}
One can choose how many components of $\Delta$ to propose per iteration of this joint update; in our applications, proposing one new component per iteration tends to admit a healthy acceptance probability between  0.3 to 0.6. 

Finally, we update the latent data $X$ using a Metropolis-Hastings step \citep{morsomme2022uniformly} conditional on $\beta$ and $\gamma$. Since the number of latent infection and recovery times can be quite large, the success of DA-MCMC hinges on the construction of an efficient proposal for the latent data. The model specification in the previous section is carefully designed to remain tractable to this end: in particular, it allows us to make use of efficient proposals using the \textit{piecewise decoupled SIR}  (PD-SIR) process. Briefly, the PD-SIR closely resembles the SIR, with the only difference being that the infection times follow a linear death process within each time interval $(c_{t-1}, c_t]$, with details in \citet{morsomme2022uniformly}. It enjoys rate-linearity, leading to closed form expressions for conditionally generating latent data that match the observed infection counts. Denoting the density of the PD-SIR by $q$, We accept $X^*$ with probability
\begin{align*}
    \alpha_2 = min\{1, \frac{L(\Delta^j, \beta^j; X^*)q(X^{(j-1)}| \Delta^j, \beta^j)}{L(\Delta^j, \beta^j;  X^{(j-1)}) q(X^{*} | (\Delta, \beta)^j)}\}
\end{align*}
That is, though the latent data are proposed using a recent technique from an approximating process, the Metropolis-Hastings correction ensures that the accepted draws correspond to the exact model posterior under the time-inhomogeneous SIR model.  

Below, we provide a pseudocode summary of the overall sampler in Algorithm \ref{alg:cap}.

\begin{algorithm}
\caption{Data-Augmented MCMC for sampling from the posterior distribution}\label{alg:cap}

\begin{algorithmic}
\Require observed data Y and initial parameters ($\Pi^{(0)}$, $\Delta^{(0)}$, $\beta(t)^{(0)}$)

\State \Return $\{(\Pi^{(j)}, \Delta^{(j)}, \beta(t)^{(j)})\}_{j=0}^N$

\State $X^{(0)} \gets draw(. \mid (\Delta, \beta)^{(0)}, Y)$ \text{ (initial latent data using the PD-SIR process) }

\For{j = 2, ..., N}

    \State \begin{varwidth}[t]{\linewidth}
    
    \textbf{1. Gibbs Update on $\Pi_{2x2}$: }

    $\pi_{01}^{(j)} \mid \Delta^{(j-1)} \sim Beta(a_{01} + n_{01}, b_{01} + n_{00})$ \par
    $\pi_{11}^{(j)} \mid \Delta^{(j-1)} \sim Beta(a_{11} + n_{11}, b_{11} + n_{10})$ 
    
    \end{varwidth}\\
    
    \State \textbf{2. Metropolis-Hastings Update on ($\Delta$, $\beta(t)$): }

    \State $\Delta^* | \Pi^{j} \sim \nu_1 \prod_{i=1}^n \pi_{\Delta_{i-1}, \Delta_i}$

    \For{k = 1, ..., K}
        \State $\beta_{k}^* | \Delta^* \sim Gamma \left( a_0 + n_{T_k}, b_0 + \int_{c_k}^{c_{k+1}} I(t)S(t) dt \right)$
    \EndFor
    

    \State Accept $(\Delta^*, \beta^*)$ with probability $\displaystyle \min\{1, \frac{L(\Delta^*, \beta^*; X)P(\beta^* \mid \Delta^*)q(\beta^{j-1} \mid \Delta^{j-1})q(\Delta^{j-1} \mid \Pi^{j})}{{L(\Delta^{j-1}, \beta^{j-1}; X)P(\beta^{j-1} \mid \Delta^{j-1})q(\beta^* \mid \Delta^*)q(\Delta^* \mid \Pi^{j})}}\}$. 
    
    
    
    \State \textbf{3. Metropolis-Hastings Update on Latent Data X: }
    
    \State $X^* \gets draw(. \mid (\Delta, \beta)^j, Y)$ \text{ (generated using the PD-SIR process) }
    

    \State Accept $X^*$ with probability $\displaystyle \min\{1, \frac{L(\Delta^j, \beta^j; X^*)q(X^{(j-1)}| \Delta^j, \beta^j)}{L(\Delta^j, \beta^j;  X^{(j-1)}) q(X^{*} | (\Delta, \beta)^j)}\}.$
    
    

\EndFor 

\end{algorithmic}
\end{algorithm}

\section{Results}
\label{sec:results}

\subsection{Simulation study}
\label{sec:results-sim}

To validate the performance of our MCMC, we begin with simulation studies on synthetic datasets. First, we simulate an epidemic from a piece-wise constant transmission rate. As a result, we produce a trajectory with clear deviations from a homogeneous SIR trajectory, which reflects the nature of the observed data in both the Ebola and COVID-19 case studies later on. Next, we demonstrate that our model is still able to produce a close, yet interpretable, piecewise approximation of the truth under a misspecified setting in which the transmission rate value varies smoothly over time.

\begin{figure}[!htp]
    \centering
    \includegraphics[clip, trim=0.6cm 4cm 0cm 3cm, width=0.75\linewidth]{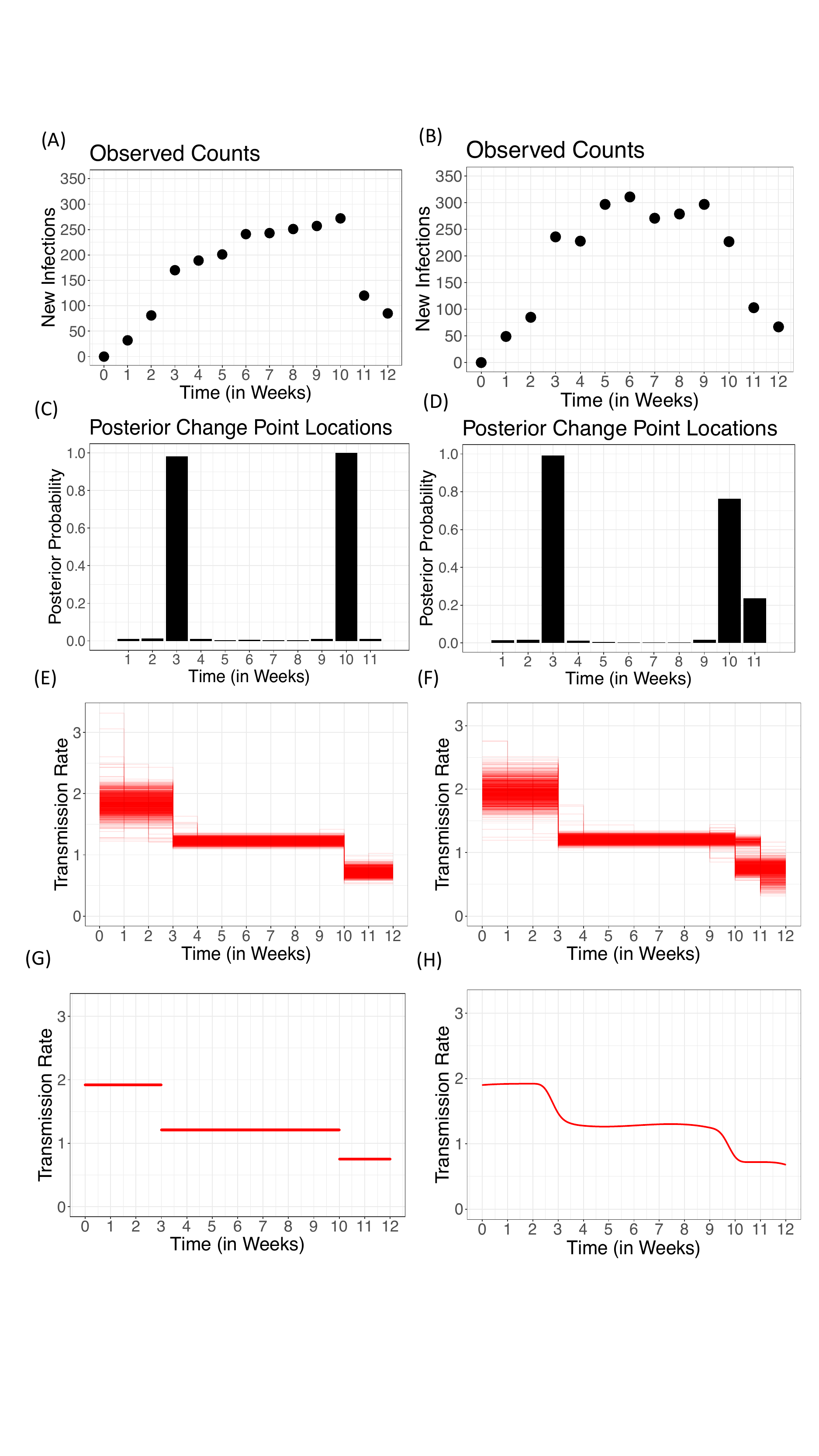}
    \caption{Simulation study results: setting 1 is displayed in the left column and setting 2 in the right column.
    (A) Simulated incidence data under setting 1. (B) Simulated incidence data under setting 2. (C) The marginal posterior probabilities of a change point occurring at each week under setting 1. (D) The marginal posterior probabilities of a change point occurring at each week under setting 2. (E) Plot of the posterior samples for $\beta(t)$ under setting 1. (F) Plot of the posterior samples for $\beta(t)$ under setting 2. (G) True piece-wise transmission rate value under setting 1. (H) True smoothly-varying transmission rate value under setting 2. 
    (note: all transmission rate values are multiplied by a factor of 10,000 for readability.) }
\label{fig:sim_results}
\end{figure}

The shape of the trajectory and placement of change points in the first simulation setting are designed to resemble the Ebola outbreak data.  
In this setting, we take  $S_0 = 10000$, $I_0 = 10$, with change points located at observation times times $3$ and $10$. The true parameter values are set at $\beta$ = (1.75e-4, 1.25e-4, 0.75e-4), and $\gamma$ = 1. We observe the outbreak at twelve observation intervals, and in the realization we considered here, the observed incidence counts are given by $\textbf{I}$ = (32, 81, 170, 189, 201, 241, 243, 251, 257, 272, 120, 85). For reference, the maximum likelihood estimate for the transmission rate, obtained under knowledge of the complete data and true change point locations, is $\hat{\beta} = (1.92e-4, 1.21e-4, 0.75e-4)$.  

The observed incidence data are shown in Figure \ref{fig:sim_results}. This trajectory displays clear deviations from the typical shape of the SIR, indicating that the homogeneous model is a poor choice, and suggesting the presence of potential change points. We chose hyperparameters to be uninformative; for each segment of $\beta$, we chose the weakly informative prior Gamma(1,1). For the prior on $\Pi$, we used Jeffrey's prior on both $\pi_{01}$ and $\pi_{11}$. We assume the recovery rate $\gamma$ is known, fixing it to the true value.

The algorithm generates samples efficiently, taking less than 10 minutes to run 50,000 iterations of the chain on a single laptop machine. More specifically, it achieves an effective sample size per second (ESS/sec) of 2.39 in this setting; results are shown in column one of Figure \ref{fig:sim_results}. Notably, even with uninformative priors on the number of change points, the algorithm is successful in assigning high probability only to the locations of true change points, empirically validating the Bayesian Occam's razor principle \citep{jefferys1992ockham}. Moreover, we observe very little bias in estimates for each segment of the step-function, with the true transmission rate values well-contained within the 95\% equal-tailed Bayesian credible intervals. This is noteworthy, since only 12 observed time points in the epidemic trajectory were supplied to the algorithm toward recovering such estimates. The posterior reflects less precision about the positioning of the second change point.

Although the Bayesian credible intervals are generally narrow, we note that the variance of the estimates were larger near the front and end of the epidemic. This is evident in the posterior draws of $\beta(t)$ shown in (Figure \ref{fig:sim_results}). This can be understood both epidemiologically and statistically: it is more difficult to conduct inference on epidemic parameters observing only the tail or early stages of an epidemic. Indeed, note that the gamma distribution describing the conditional for $\beta_k$ is $(a_0 + n_T) / (b_0 + \int_{c_k}^{c_{k+1}} I(t)S(t) dt)^2$; the term $I(t)$ appearing in the denominator tends to be small at the beginning of an outbreak or at its end, resulting in wider credible intervals. This perspective is unavailable in the time-homogeneous case when inference is based around one constant value $\beta$ that describes the entire outbreak.

To evaluate the convergence properties of the Markov chain, we run 3 separate chains with over-dispersed starting values $\beta_0$ = ($0.1 \cdot \beta$, $\beta$, $10 \cdot \beta$). Despite being initialized in various low-density regions, each chain appears to converge as evident from both the trace plots and a multivariate Gelman-Rubin convergence diagnostic of less than 1.01 (Figure \ref{fig:sim1_diagnostics}). The frequency of flips between ones and zeros for each $\{\Delta_k\}$ is a good indication that the MCMC mixes well and explores with high frequency the space of candidate change point configurations \ref{fig:sim1_diagnostics}.

In reality, the drops in transmission rate following a policy intervention may not be abrupt. To test the performance of our model in this setting, we next run our model on an epidemic that was generated by a smoothly-varying, rather than piece-wise, transmission rate. Specifically, this smoothly-varying rate was generated using a cubic B-spline with cut-points at times 2, 2.5, 3, 3.5, 4, 9, 9.5, 10, 10.5, and 11. Keeping the same initial conditions, $S_0 = 10000$ and $I_0 = 10$, this new transmission rate resulted in a trajectory with observed counts $\textbf{I}$ = (49, 85, 236, 228, 297, 311, 271, 279, 297, 227, 103,  67) (Figure \ref{fig:sim_results}b). As in the first simulation setting, we keep hyperparameters uninformative, and assumed that the recovery rate $\gamma$ is known, fixing it at the true value.

In contrast to in the piece-wise setting, in which both change points produced visibly notable deviations from the time-homogeneous SIR curve in the observed infection counts (Figure \ref{fig:sim_results}a), the continuous transmission rate presented a visibly more challenging problem. Specifically, the observed case count data of Figure \ref{fig:sim_results}b, show a less visible deviation from the typical time-homogeneous SIR, indicating that the homogeneous model may be a reasonable model under this observed trajectory. In particular, the more gradual decline in transmission rate around weeks nine through eleven (shown in Figure \ref{fig:sim_results}h) are less abrupt. However, the sampler still suggests with high posterior probability a drop in transmission rate around weeks ten through eleven, show that our model is able to give an interpretable, yet close, piece-wise approximation of $\beta(t)$ (Figure \ref{fig:sim_results}f). 

\subsection{Comparison to existing methods}

\begin{figure}[!ht]
\centering
    \includegraphics[clip, trim=4cm 3.5cm 5cm 3.5cm, width=0.8\linewidth]{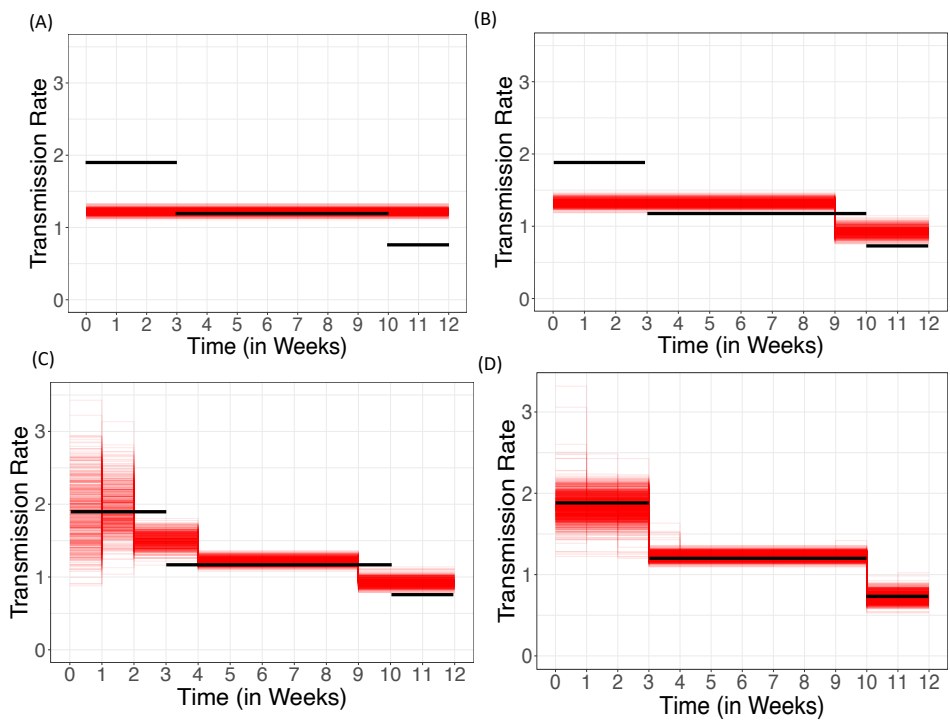}
    \caption{(A) Samples from the posterior transmission rate over time, recovered from the time-homogeneous model. (B) Samples from the posterior transmission rate over time, recovered from the fixed change point model. (C) Samples of the posterior transmission rate over time, recovered using Binary Segmentation and PELT. (D) Samples from the posterior transmission rate over time, recovered using our proposed model. Black lines indicate the true transmission rate values. 
    }
    \label{fig:comparison_models}
    \end{figure}

We next compare the proposed model to the time-homogeneous stochastic SIR model described in \citep{morsomme2022uniformly}, the fixed change point models described in \cite{lekone2006statistical} and \cite{ward2021individual}, and popular general-purpose change point detection algorithms which ignore SIR dynamics \citep{fearnhead2019changepoint}. All of these previous approaches resulted in biased estimates for $\beta(t)$ in at least one segment (Figure \ref{fig:comparison_models}). The time-homogeneous model's estimate for $\beta$ fails to capture the presence of change points in the data, resulting in a biased estimate that is the average of three notably disparate regimes (Figure \ref{fig:comparison_models}). A fixed change point model is a model that assumes the changes in transmission occur at known locations, frequently assumed to be the exact time of intervention. This is frequently done in practice in the epidemic modeling literature \citep{lekone2006statistical} \citep{ward2021individual}. Although these rigid change point models allow for a piece-wise transmission rate, they lead to biased estimates of the effects of control interventions by assuming, rather than learning, the change point locations. Of course these models fail to account for additional change points that are not known or specified by the user, resulting in biased estimates of the magnitudes of the drops (Figure \ref{fig:comparison_models}b). Finally, to show the inaccuracies that arise when we disregard the SIR dynamics altogether, we compare our method to two classical change point detection algorithms, Binary segmentation and PELT \citep{fearnhead2019changepoint}. Both algorithms were designed to detect change points in the rate of a Poisson model. However, these two methods do not benefit from the additional information given by SIR compartment sizes and dynamics toward detecting change points, resulting in biased estimates of the transmission rate (Figure \ref{fig:comparison_models}c). This illustrates the advantage of learning change points of the transmission rate function, over casting it in a classical setting operating on the count data themselves in a way that is model agnostic.  Within these comparisons, we see that our proposed method is advantageous for detecting change points in the context of epidemic count data.

\subsection{Applications to outbreaks of Ebola and COVID-19}

We next consider case studies on data collected from the Ebola outbreak in Guinea in 2014, and a COVID-19 outbreak on a university campus. 

\begin{figure}[!ht]
    \centering
    \includegraphics[clip, trim=5cm 0cm 10.5cm 0cm, width=0.80\linewidth]{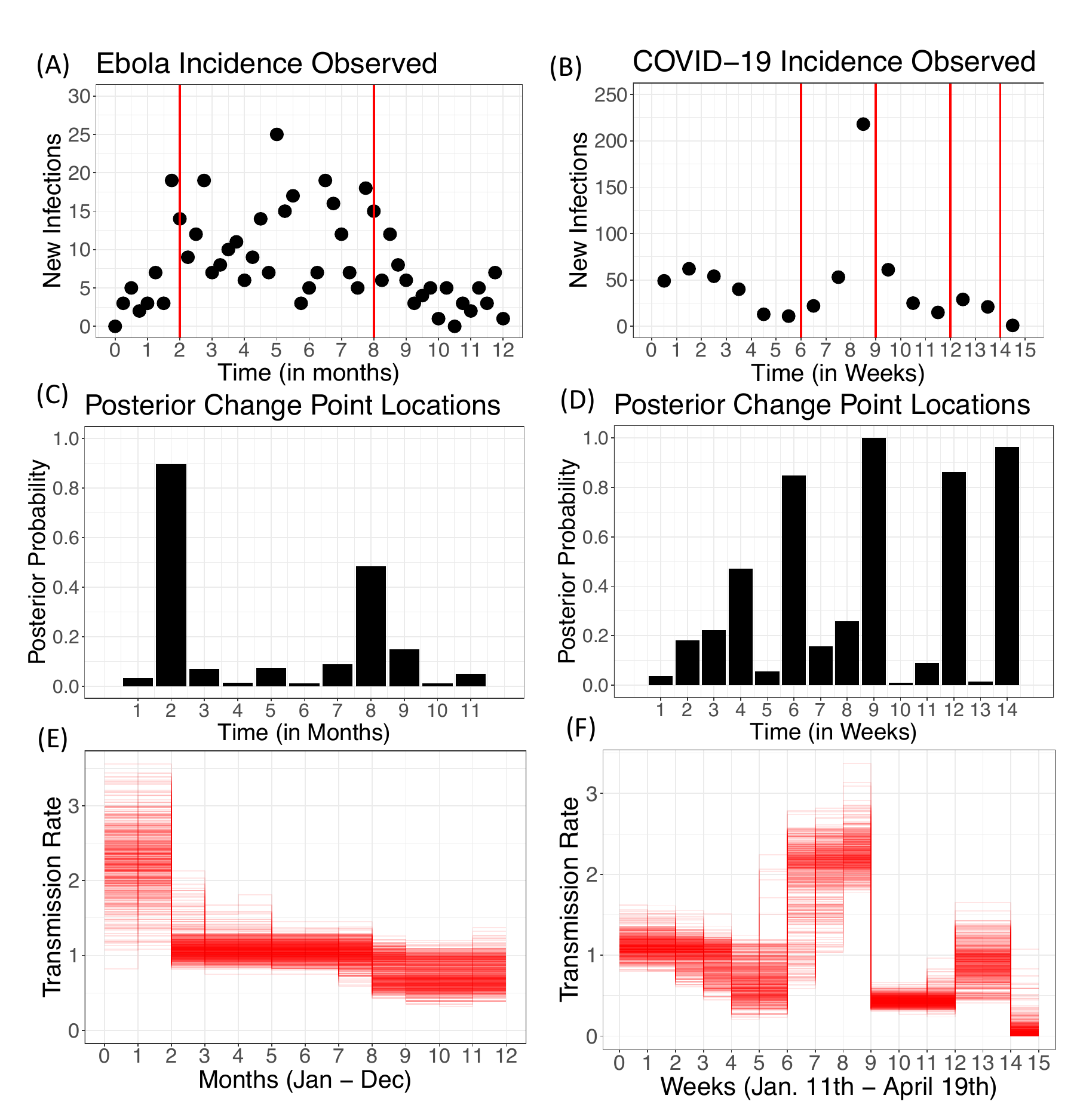}
\caption{Application to two real outbreaks: Ebola and COVID-19. (A) Counts of weekly new infections of Ebola cases in Guéckédou, Guinea from January to December, 2014. Red lines mark the positions with high posterior probability of being a change point.
(B) Counts of weekly new infections of Covid-19. Red lines indicate positions with high posterior probability of being a change point.
(C) The marginal posterior probabilities of a change point occurring at each month for the Ebola outbreak.
(D) The marginal posterior probabilities of a change point occurring at each week for the COVID-19 outbreak.
(E) Plot of the posterior samples for the transmission rate for Ebola in Guéckédou, Guinea. 
(F) Plot of the posterior samples for the transmission rate for COVID-19. 
(all transmission rate values are scaled by the population size for readability.)
}
\label{fig:application_to_real_data}
\end{figure}

The Ebola dataset we study consists of weekly infection counts from January 1st to December 31st, 2014, collected in Guéckédou, Guinea, a prefecture with 292,000 inhabitants. These data were collected over a time period in which control measures where implemented at various times \citep{coltart2017ebola}, making it a natural candidate for modeling the force of infection to vary over time. Applying our model with piece-wise components at a monthly resolution, we obtain results displayed in the left column of Figure \ref{fig:application_to_real_data}. Figure \ref{fig:application_to_real_data}a shows the observed cases of Ebola reported over time on a weekly basis, with the red lines marking places of the most likely change point occurrences, according to the model. The model suggests the most likely change point locations to be in March and August. Figure \ref{fig:application_to_real_data}c shows the marginal posterior probabilities of a change point occurring at each month. Figure \ref{fig:application_to_real_data}e shows a plot of the posterior samples of transmission rate.

The change points in the data may give insights into the efficacy of the control measures implemented in Guinea in 2014. Interpreting these results, we note that March marked the closing of public schools in Guéckédou and the confirmation of Ebola as the infectious agent \citep{coltart2017ebola}. August marked a closing of borders with Sierra Leone and Liberia and the implementation of exit screening and flight cancellations. Our findings in terms of changes in $\beta(t)$ coinciding with these measures suggest they are associated in slowing the spread of the outbreak, as reflected in a lower inferred force of transmission. An important quantity in the spread of an epidemic is the effective reproduction number $R(t)=\frac{\beta(t)}{\gamma}S(t)$ which corresponds to the expected number of secondary infections arising from one infectious individual at time $t$. The posterior trajectories of $R(t)$ drop in value from 2.25 to 1.08, and then then to 0.74 following August. These results indicate that the virus was initially spreading rapidly in the population, $R(t)>1$, before being brought under control at the end of the year, $R(t)<1$.
Our model did not, however, pick up a change in October when the government banned country-wide celebrations of Eid \citep{coltart2017ebola}, suggesting that this measure may have had a less significant effects on the transmission rate as reflected in the data.

We next consider inference using data from Duke University's COVID-19 Testing Tracker in the spring of 2021. The data consist of weekly new positive case counts for students, faculty, and staff across campus from January 11th to April 25th, 2021. During this time, Duke conducted sample testing without respect to reported symptoms. At the end of week 9, the university implemented a stay-in-place order (March 8-14th), during which students were required to stay in their residences at all times except for essential activities related to food, health, or safety. Figure \ref{fig:application_to_real_data} shows the results of our model applied to these incidence data.

Figure \ref{fig:application_to_real_data}b shows the observed cases of COVID-19 reported over time on a weekly basis, with the red lines marking all places of likely change point occurrences, according to the model. Figure \ref{fig:application_to_real_data}d shows the marginal posterior probabilities of a change point occurring at each month. Figure \ref{fig:application_to_real_data}f shows a plot of the posterior samples of transmission rate.
Here, our method picks up a pronounced change, where we observe a 4.5 fold decrease in $\beta$ from week 9 to week 10, suggesting that the university-wide stay-in-place order was highly effective in reducing the spread of the virus (Figure \ref{fig:application_to_real_data}) 
Moreover, our model estimates that $R(t)$ dropped below 1 following the stay-in-place order, suggesting that the spread of the virus was at least temporarily under control following the mitigation policy.

\paragraph{Model diagnostics}

\begin{figure}[!htp]
    \centering
    \includegraphics[clip, trim=0.6cm 2cm 0cm 4cm, width=0.9\linewidth]{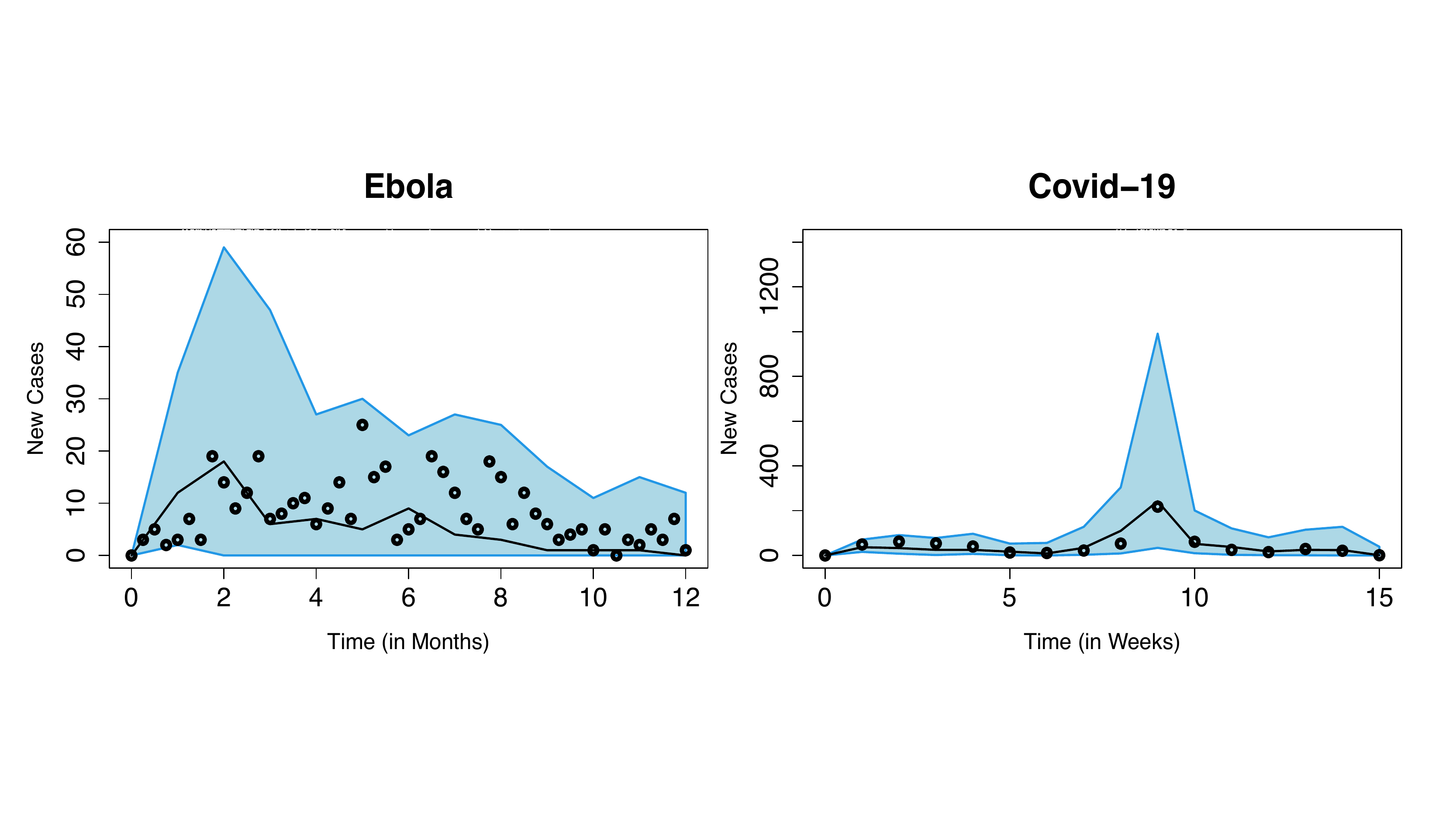}
    \caption{Posterior predictive distributions of new cases with 95\% posterior credible interval plotted in blue; the posterior mean is marked by the black line, with raw observed data plotted as black dots.}
\label{fig:posterior_predictive_checks}
\end{figure}

To assess model fit, we perform posterior predictive model checks based on the number of new cases over time. For each data set, we simulate epidemic trajectories from the posterior predictive distribution and then compute the trajectory of new cases over time in these simulations. This produces a posterior predictive distribution of cases over time, as plotted in Figure \ref{fig:posterior_predictive_checks}. For both the Ebola and COVID-19 case studies, the observed data fall reasonably within the 95\% posterior predictive interval. We note that there is larger variation in the stochastic trajectories simulated by the COVID-19 model. This is in line with what one would expect, given that the observed data consist of only 15 observation times in the latter case study.

\section{Discussion}

\label{sec:discussion}
This article develops a novel framework for exact Bayesian inference in a stochastic SIR model with a piece-wise constant time-varying transmission rate $\beta(t)$.  The model is designed to be flexible enough to capture time-heterogeneity in the infection rate while remaining amenable to  sampling targeting the exact model posterior. The resulting method yields interpretable parameter estimates favoring model parsimony and comes equipped with corresponding uncertainty estimates. In particular, we propose a data-augmented MCMC algorithm that samples from the joint posterior distribution of the model parameters, performing change point estimation in the process. The framework is user-friendly and allows for prior specification on how strongly one favors constant runs in the transmission rate,  leveraging recent developments in efficient proposals for the latent space of unobserved epidemic events when the observed data consist of incidence counts.

Detecting changes in disease spread allows us to gain valuable insight into the efficacy of mitigation policies that were implemented, informing us on where and how quickly a specific policy begins to be effective. While we show that the method is successful in assigning high probability to the true change points in simulations where the ground truth $\beta(t)$ is piecewise constant, our method provides interpretable and flexible approximate estimates even when the ground truth transmission rate is continuously varying. From a practical perspective, the posterior plots of the learned transmission rate are easy to interpret for non-statisticians; not only can one immediately see the uncertainty of each segment, but the most pronounced change points as supported by the data are visually obvious. Finally, because we have a continuous-time generative model, the sampler is ready to be used off-the-shelf for incidence data even when the observation intervals are irregular.

The methodology proposed here leads naturally to several extensions. We have seen success with the piece-wise model in empirical studies in this paper, and future work may analyze recovery guarantees from a theoretical perspective. The focus of this paper is on developing the model framework and enabling efficient Bayesian computation; moreover, many interesting use cases of the method extend to settings where the true transmission rate need not be piecewise constant. Not only does this change point model lend a natural interpretation for getting a direct quantity of the drop, or decline, in transmission rate following a policy intervention, it is also able to segment the epidemic into different stages based on transmission rate. Along these lines, future studies may aim to model a more flexible transmission rate, incorporating time trends such as seasonality in terms beyond piecewise constant building blocks. Past work using more flexible functions such as Gaussian processes have been successful, though they incur significantly higher computational cost and do not scale to large outbreaks due to operations such as matrix inversions \citep{kypraios2018bayesian} \citep{xu2016bayesian}. In some cases, these samplers must run for hours for small outbreaks with only a hundred infections \citep{xu2016bayesian}. Our proposed framework can generate thousands of posterior samples for outbreaks with hundreds of infections in just several minutes on a single machine, suggesting there is room for further model complexity while retaining computational tractability. Additionally, in settings where more information about the outbreak is available, one may consider defining a regression model for $\beta(t)$ based on covariates related to control measures or models of human behavior \citep{mcadams2021equilibrium}. Levels of mobility, levels of vaccination, and other auxiliary indicators of transmission \citep{mcdonald2021can} may further be incorporated into the rate function. The broader inferential framework proposed here readily extends  to such settings.

While the sampler scales efficiently in per iteration computational cost, allowing applications to much larger outbreaks, our empirical studies have focused  on the setting where the number of data points is small.

In this setting, our DA-MCMC algorithm mixes adequately, as evident in the trace plots for $\Delta$ in Figure \ref{fig:sim1_diagnostics}, which exhibit frequent jumps between $0$ and $1$.
The space of possible change points is immense and it is well known to be an extremely difficult problem to efficiently sample over this space.
Given this it is as expected that our MCMC takes a longer time to mix for select parameters that border a change point. For these $\beta_k$'s, the posterior distribution can appear bimodal, with the two modes corresponding to two different piece-wise regimes. To further improve mixing for these parameters, future work may design informed proposals not only based on the structure of the disease model but also on the discrete space of change point configurations. One fruitful line of thought along these lines for high-dimensional discrete problems is the locally balanced proposal described in \cite{zanella2020informed}, which uses point-wise informed proposals and balancing functions to bias proposals toward high-probability states. 

Our data augmentation approach complements classical change point methodology, but future work may seek to further integrate these powerful ideas within  stochastic epidemic models. For instance, \cite{fearnhead2006exact} and \cite{fearnhead2019changepoint} proposed a set of dynamic programming algorithms to perform direct simulation from the posterior distribution of change point positions. These dynamic programming approaches minimize the cost over all possible segmentations of the data exactly under a given statistical criteria, sidestepping the need to design careful MCMC moves when there is a large space of models to explore. However, to ensure the necessary Markov property that is required for the forward-backward recursions required by these algorithms, the methods require that the observations are conditionally independent on the change points and the parameter values --- an assumption that does not hold for SIR, case counts explicitly depend on past values of $S,I$. Similarly, \cite{cappello2021scalable} propose a Bayesian variable selection procedure for multiple change point detection  not requiring MCMC. Again, these methods are developed in the context of detecting changes in mean under Gaussian observations, for which they designed a sequential procedure for deriving marginal posterior probabilities of change points in closed form. Incorporating these methodologies toward parametric change-point detection within mechanistic models such as the SIR remains a promising future direction in a challenging dependent data setting.

\section{Implementation} \texttt{R} code is available at \url{ https://github.com/JennyHuang19/damcmc_timevarying_beta}, including all  open-source code and scripts for reproducing the results in this paper. 

\section{Acknowledgments}

This work was partially supported by NIH grants R01ES028804 and R01ES027498, and NSF grants DMS-2230074 and PIPP-2200047. We thank the 2021 DOMath team (Trevor Bowman, Gregory Orme, Pranay Pherwani, Omar Melikechi, and Tao Tang) for contributing valuable conversations in the early stages of this project.

\bibliography{references}  

\section{Supplemental Material}

\subsection{Choice of Priors on Change Point Locations}

\begin{figure}[!t]
\centering
    \includegraphics[clip, trim=0cm 3.5cm 0cm 3.5cm, width=0.9\linewidth]{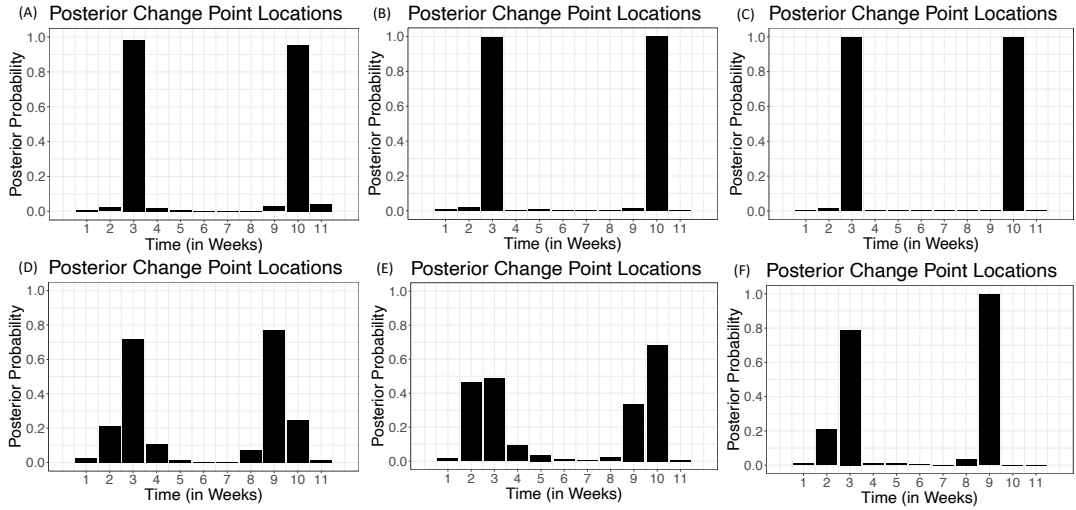}
    \caption{Sensitivity analysis on the choice of prior on the change point probabilities. The top row shows results under simulation setting 1 and the bottom row under setting 2. For each setting, we fix the prior on $\pi_{11}$ to be Be(1, 10), and vary the prior on $\pi_{01}$ to be:
    (A and D) Be(0.5, 0.5), (B and E) Be(1, 5), and (C and F) Be(5, 50).}
    \label{fig:change_point_sensitivity}
    \end{figure}

We performed a prior sensitivity analysis exploring the influence of the prior on $\Pi$ on the recovery of change points (Figure \ref{fig:change_point_sensitivity}). We consider the two simulated data sets from Section \ref{sec:results-sim}: the setting where transmission rate evolves as a piece-wise function and the setting where transmission rate evolves as a continuous function. The first data have a relatively high signal to noise ratio. As a result, we are able to identify the correct locations of change points with high probability using a non-informative, Jeffrey's prior, on $\Pi$. Hence, we observe that the effect of the prior on the results is negligible as the three priors for $\pi_{01}$ (Be(0.5, 0.5), Be(1, 5), and Be(5,50)) yield similar results. In contrast, the second data have more subtle changes in the infection rate. In this setting, the prior impacts the marginal posterior inclusion probabilities $\Delta$. That is, we see a reduction in false signal and an increased probability of recovering the true change point locations moving from left to right in row 2 of Figure \ref{fig:change_point_sensitivity}. 

\subsection{Convergence Diagnostics}

\begin{figure}[!hp]
    \centering

    \begin{subfigure}{0.9\linewidth}
        \includegraphics[clip, trim=0cm 1cm 0cm 0cm, width=\linewidth]{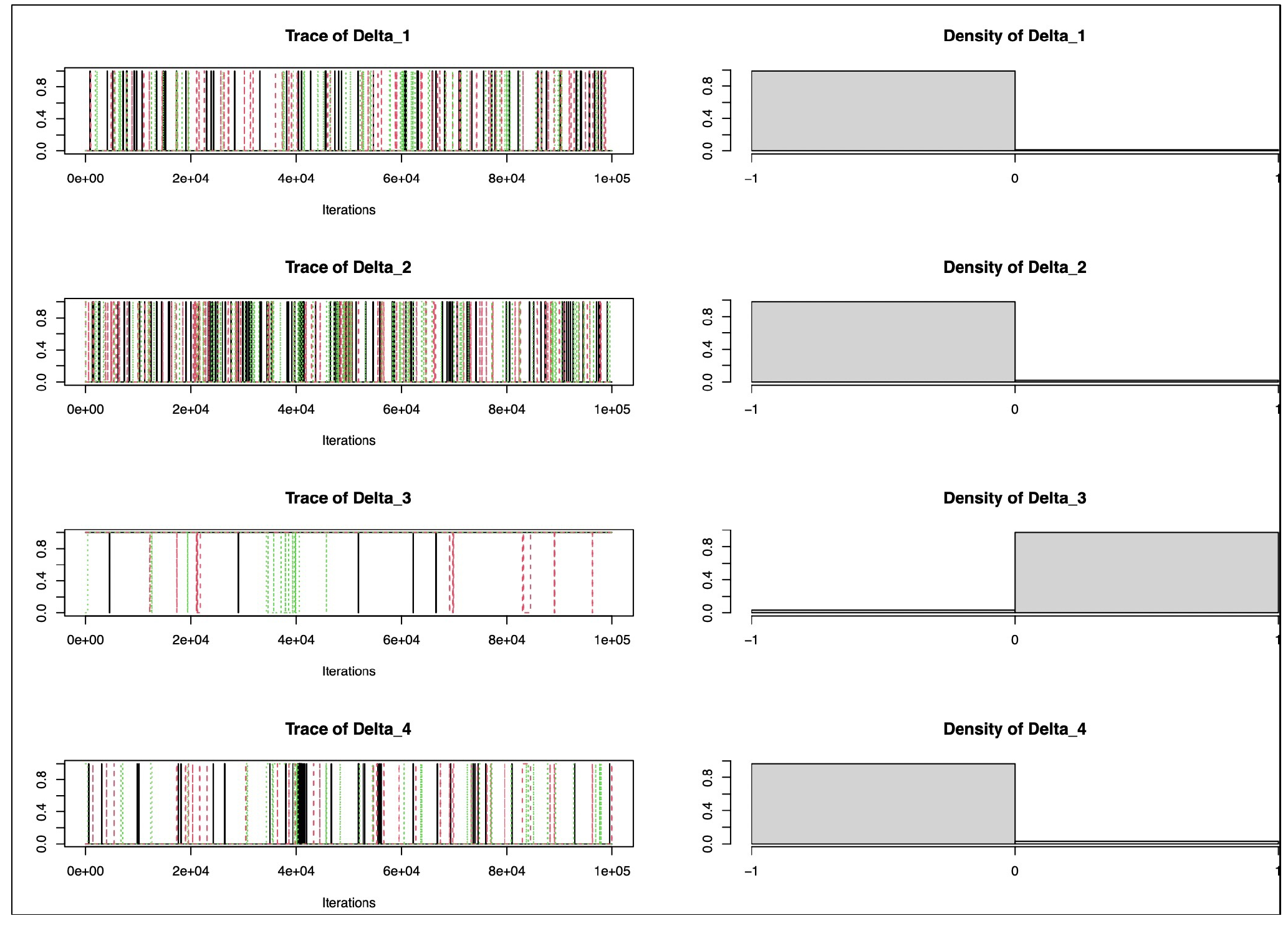} 
    \end{subfigure}

    \begin{subfigure}{0.9\linewidth}
        \includegraphics[clip, trim=0cm 0cm 0cm 0cm, width=\linewidth]{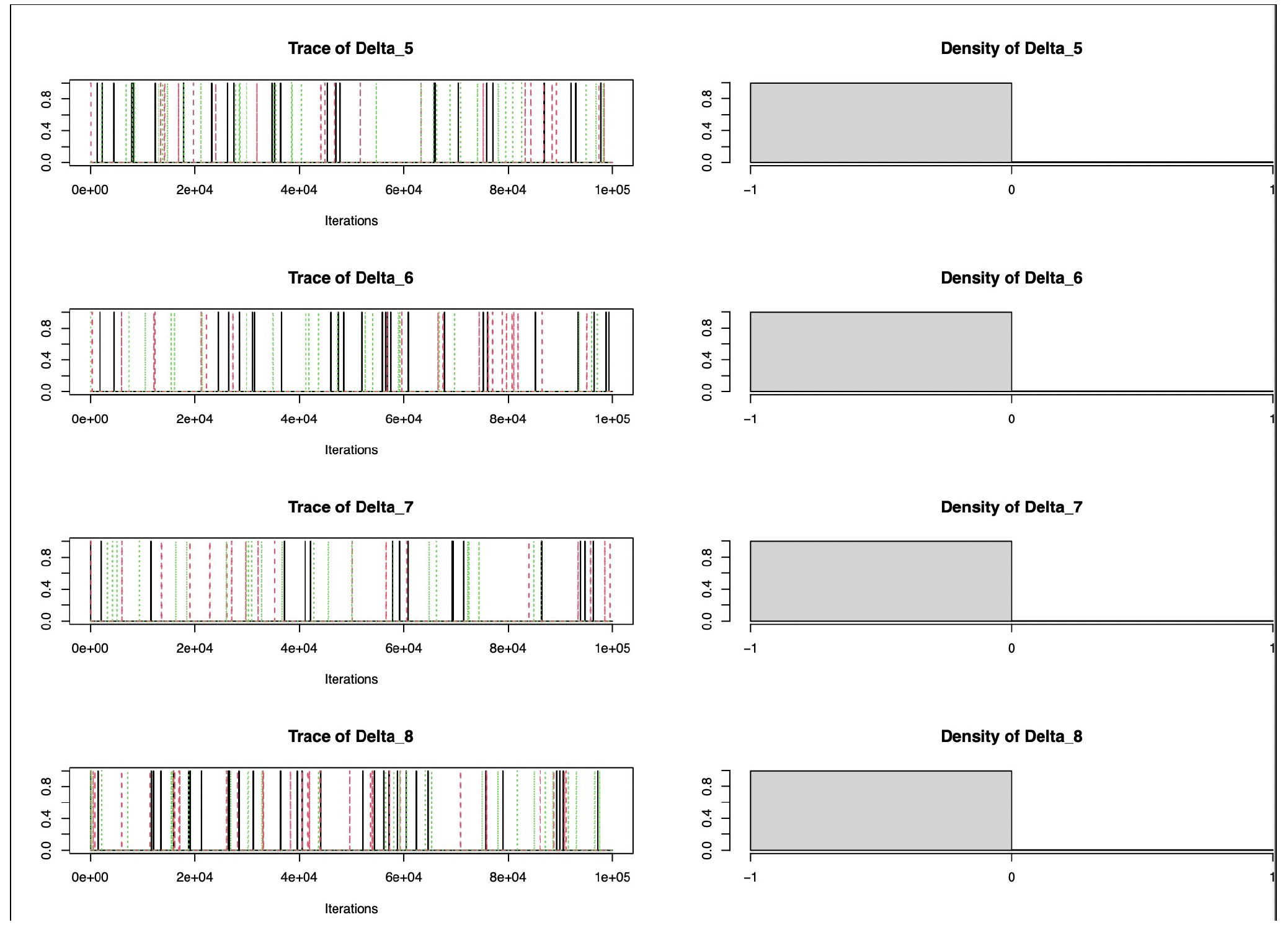}
    \end{subfigure}
    
    \caption{Simulation 1: Traceplots and marginal posterior distributions of the parameters $\Delta$. We run 3 separate chains for 100,000 iterations.}%
\end{figure}

\begin{figure}[!hp]\ContinuedFloat
    \centering
    
    \begin{subfigure}{\linewidth}
        \includegraphics[clip, trim=0cm 1cm 0cm 0cm, width=\linewidth]{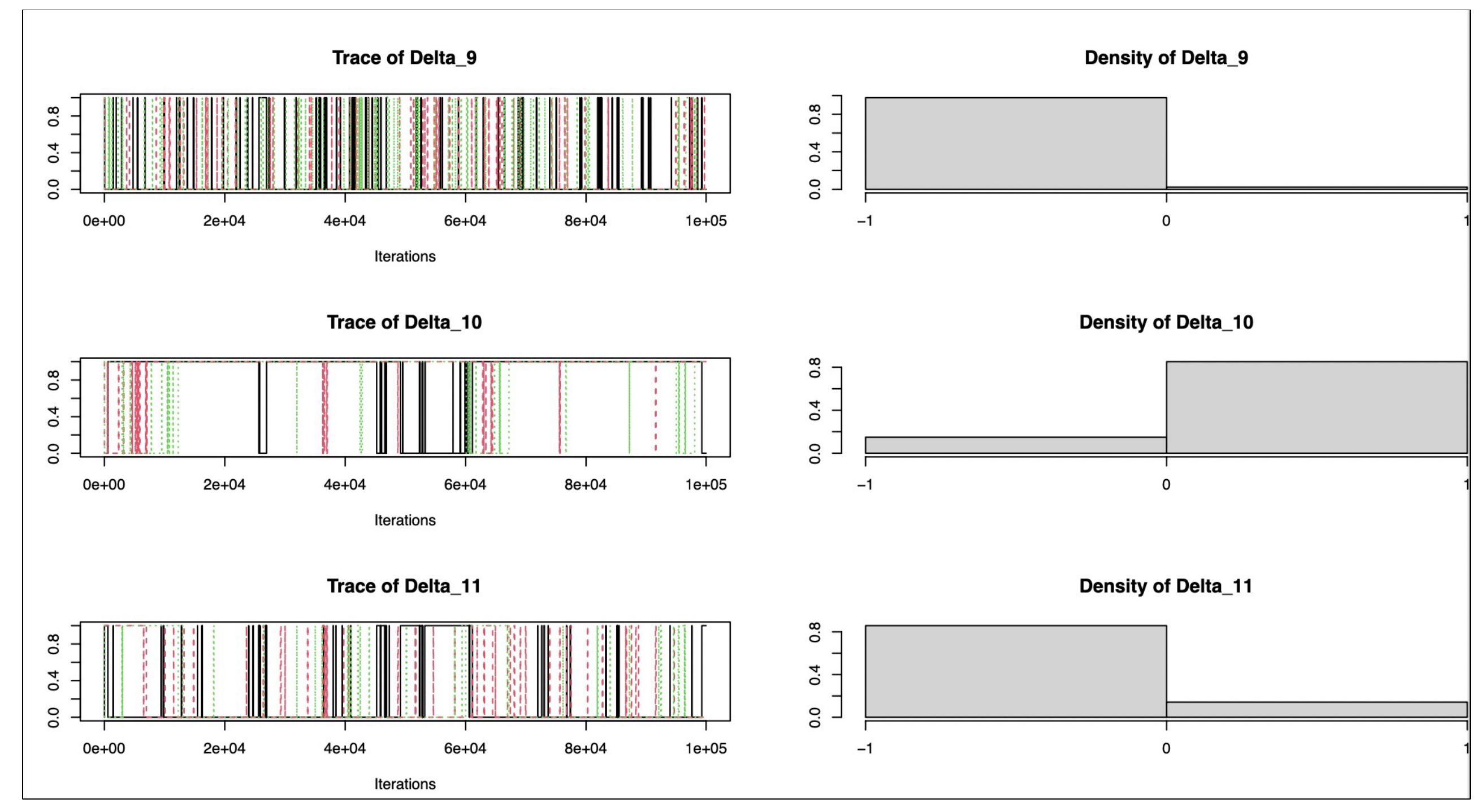} 
    \end{subfigure}
    
    \caption{(Continued) Simulation 1: Traceplots and marginal posterior distributions of the parameters $\Delta$. We run 3 separate chains for 100,000 iterations.}%
    
    \label{fig:sim1_diagnostics}
\end{figure}

\begin{figure}[!hp]
    \centering
    
    \begin{subfigure}{\linewidth}
        \includegraphics[clip, trim=0cm 0cm 0cm 0cm, width=\linewidth]{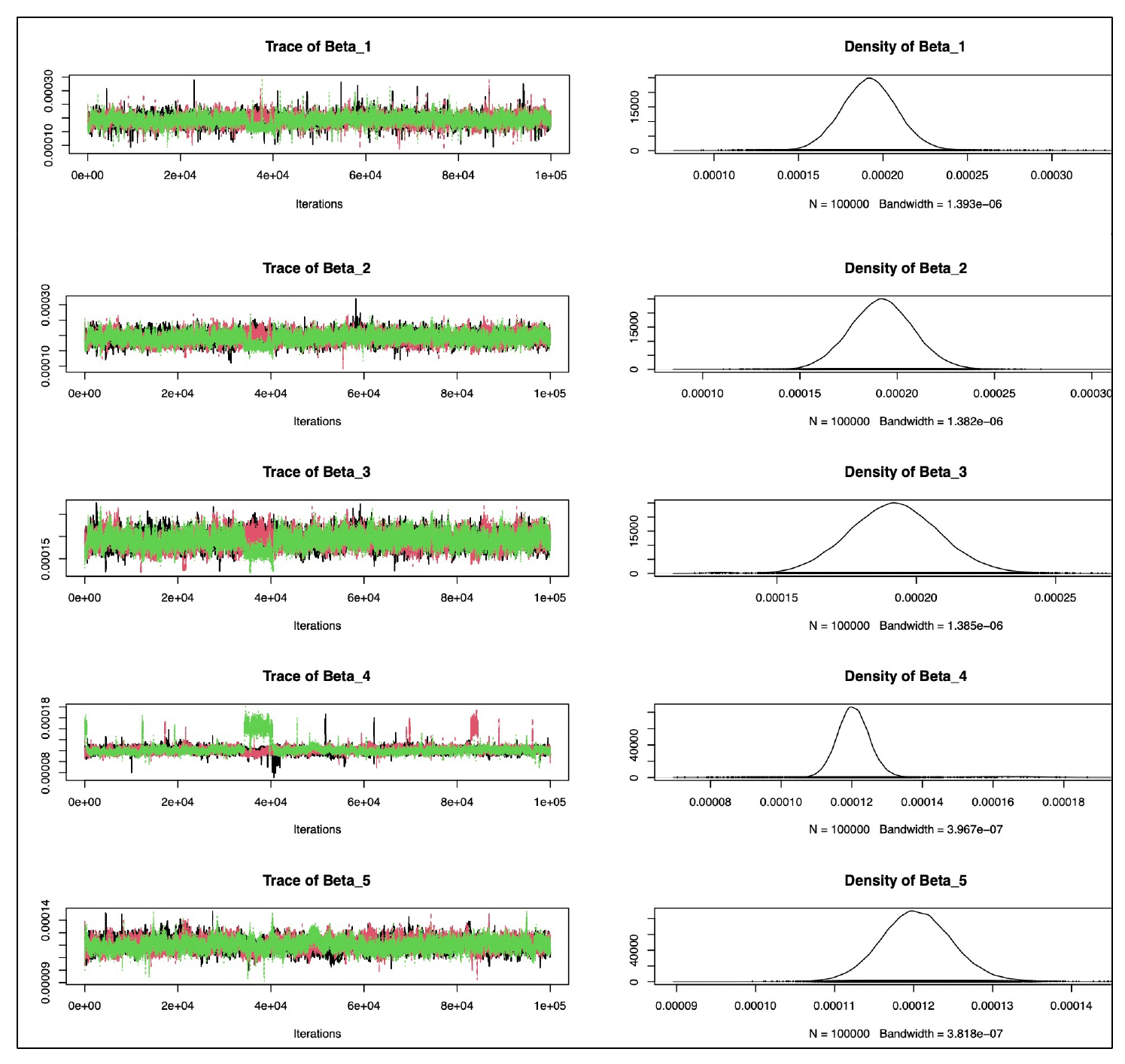}
    \end{subfigure}
    
    \caption{Simulation 1: Traceplots and marginal posterior distributions of the parameters $\beta$. We run 3 separate chains for 100,000 iterations with over-dispersed starting values $\beta_0$ = ($0.1 \cdot \beta$, $\beta$, $10 \cdot \beta$).}%
\end{figure}

\begin{figure}[hp!]\ContinuedFloat
    \centering

    \begin{subfigure}{\linewidth}
        \includegraphics[clip, trim=0cm 0cm 0cm 0cm, width=\linewidth]{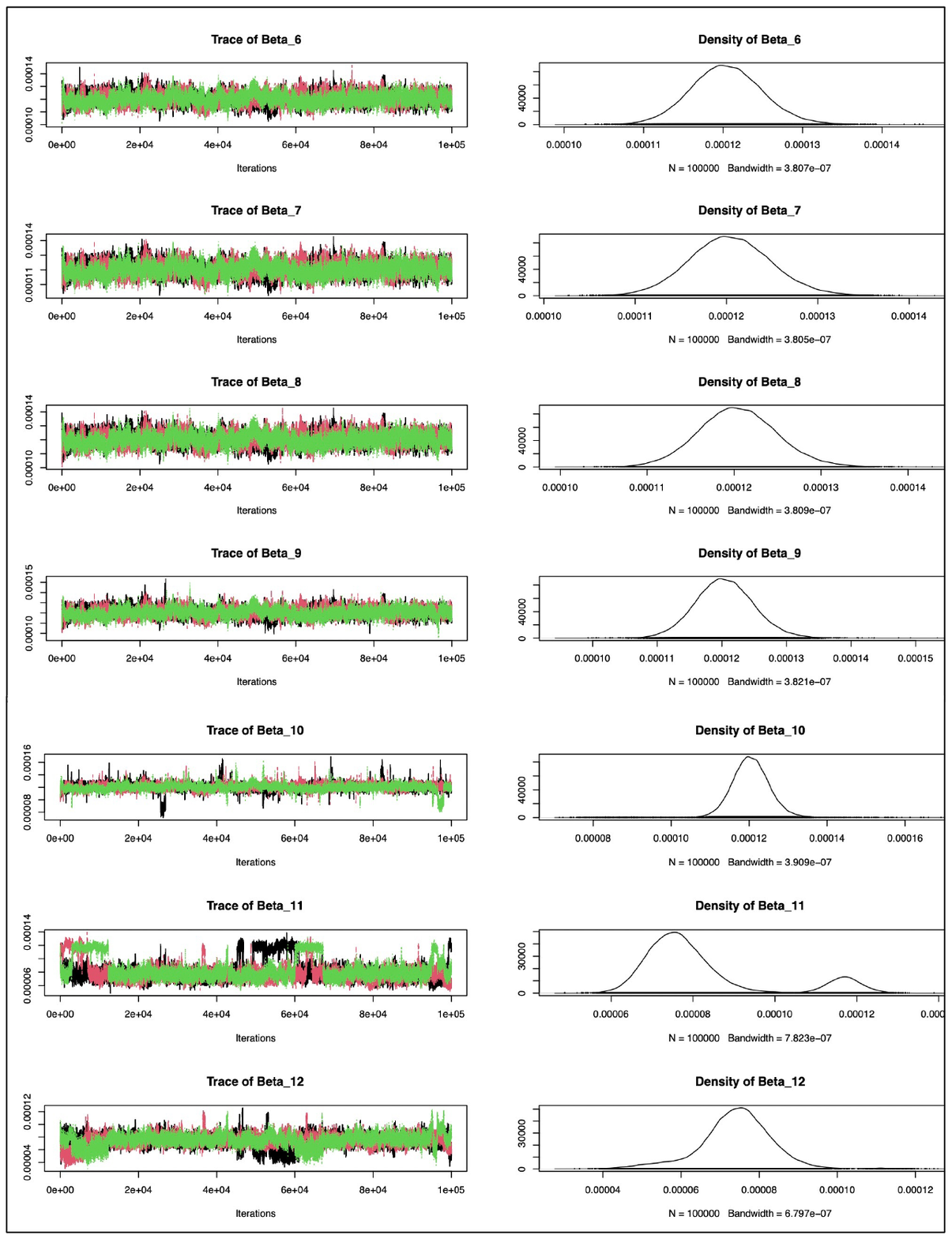}
    \end{subfigure}
    
    \caption{(Continued) Simulation 1: Traceplots and marginal posterior distributions of the parameters $\beta$. We run 3 separate chains for 100,000 iterations with over-dispersed starting values $\beta_0$ = ($0.1 \cdot \beta$, $\beta$, $10 \cdot \beta$).}%
\end{figure}

\begin{figure}[hp!]
    \centering

    {\includegraphics[width=0.8\linewidth]{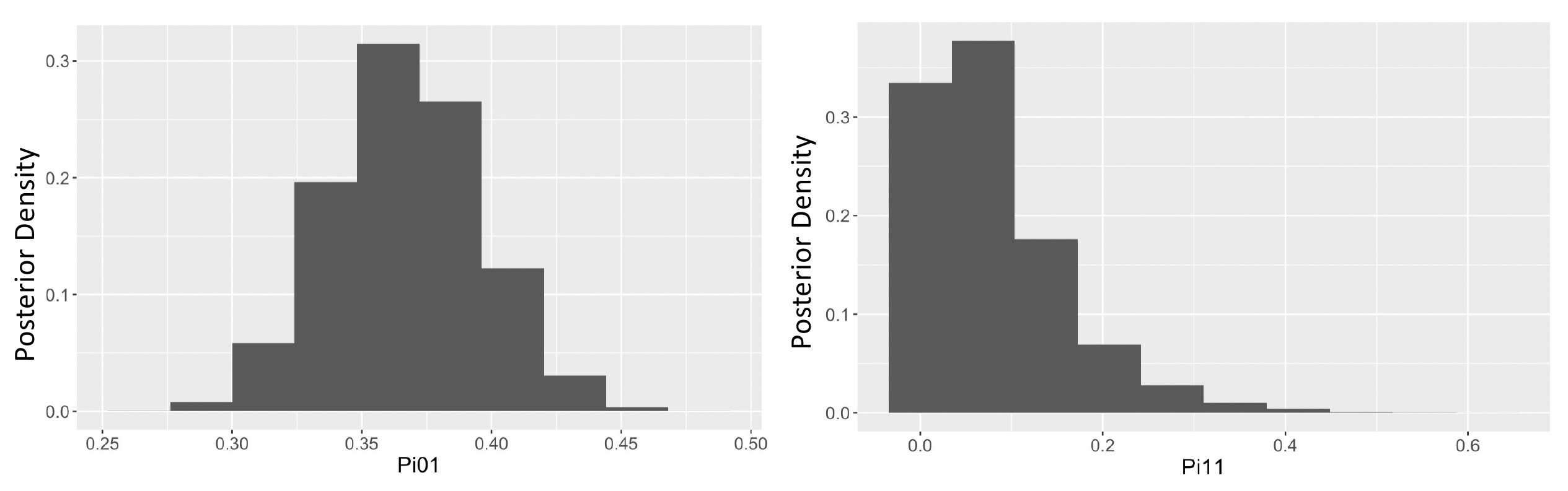}}
    
    \caption{Simulation 1: marginal posterior distributions of the parameters $\pi_{01}$ and $\pi_{11}$. We run 3 separate chains for 100,000 iterations.}%
\end{figure}

\end{document}